\documentclass[conference,10pt, a4paper]{IEEEtran}
\usepackage{comment}
\usepackage{algorithm}
\usepackage{algpseudocode}

\usepackage{amssymb}
\usepackage{amsmath} 

\usepackage{array}
\usepackage{graphicx}
\usepackage{epstopdf}
\usepackage{stmaryrd}
\usepackage{dsfont}
\usepackage{subfig}
\usepackage{paralist}
\usepackage{subfig}
\usepackage{booktabs}
\usepackage{tabularx}
\usepackage{colortbl}
\usepackage{color}
\usepackage{hhline}
\usepackage{multirow} 
\usepackage[table]{xcolor}
\usepackage{todonotes}
\usepackage{url}
\usepackage{balance}

\makeatletter
\renewcommand\paragraph{\@startsection{paragraph}{4}{\z@}%
    {1.5ex plus .2ex minus .3ex}%
            {-0em}%
                        {\normalsize\bf}}
\makeatother

\title{ARABIS: an Asynchronous Acoustic Indoor Positioning System for Mobile Devices}
\author{\IEEEauthorblockN{Yu-Ting Wang$^\dag$, Jun Li$^\ddag$, Rong Zheng$^\ddag$, Dongmei Zhao$^\dag$}
\IEEEauthorblockA{$^\dag$Department of Electrical and Computer Engineering\\
$^\ddag$Department of Computing and Software \\
McMaster University\\
Hamilton, ON, Canada\\
Email: \{{\it wangy198,lij203,rzheng,dzhao}\}@mcmaster.ca}
}

\begin{document}

\maketitle
\begin{abstract}
Acoustic ranging based indoor positioning solutions have the advantage of higher ranging accuracy and better compatibility with commercial-off-the-self consumer devices. However, similar to other time-domain based approaches using Time-of-Arrival and Time-Difference-of-Arrival, they suffer from performance degradation in presence of multi-path propagation and low received signal-to-noise ratio (SNR) in indoor environments. In this paper, we improve upon our previous work on asynchronous acoustic indoor positioning and develop ARABIS, a robust and low-cost acoustic positioning system (IPS) for mobile devices. We develop a low-cost acoustic board custom-designed to support large operational ranges and extensibility. To mitigate the effects of low SNR and multi-path propagation, we devise a robust algorithm that iteratively removes possible outliers by taking advantage of redundant TDoA estimates. Experiments have been carried in two testbeds of sizes $10.67m \times 7.76m$ and $15m \times 15m$, one in an academic building and one in a convention center. The proposed system achieves average and 95\% quantile localization errors of 7.4cm and 16.0cm in the first testbed with 8 anchor nodes and average and 95\% quantile localization errors of 20.4cm and 40.0cm in the second testbed with 4 anchor nodes only. 
%
\end{abstract}
\section{Introduction}
Location-based services (LBS) have experienced substantial growth in the last
decade with the proliferation of smart devices. A recent survey by Pew
Research Center's Internet Project found that 90\% of the US adult smartphone owners
aged 18 and older use their phones to get directions or other information based
on their current location~\cite{emarketer}. Service providers can benefit from users' location information in facilitating precision advertising, personalized
recommendation, resource tracking, proximity notification, etc.

To provide LBS, location awareness is an essential step. To date,
indoor positioning systems (IPSs) only have limited success due to low accuracy and/or high cost of infrastructure supports.
Various signal sources have been considered in infrastructure-based IPS
solutions. Most prominently, time-of-arrival (ToA) or time-difference-of-arrival
(TDoA) estimates from radio-frequency transmitters (e.g., access points,
ultra-wide band beacon nodes) have been utilized to determine the ranges or
pseudo-ranges from anchor nodes to target devices and to infer
locations of the latter~\cite{mariakakis2014sail,xiong2014synchronicity,leng2012gps}. In contrast to RF-based solutions, acoustic indoor localization has the advantage of less stringent requirements on timing accuracy and the potential to work with commercial-off-the-shelf mobile phone devices as irrespective of hardware capabilities, all mobile phones are equipped with at least one speaker and one microphone. Recently, several acoustic IPS solutions have been developed for mobile phones, including those operating in audible frequency ranges~\cite{peng2007beepbeep, xu2011whistle}, and those in human inaudible but with the hearing range of majority of smart phone devices~\cite{lazik2012indoor, lazik2015ultrasonic,wang16euc}. 

We consider indoor localization using fixed acoustic anchor nodes
that transmit acoustic signals in the human inaudible range but decodable by
smart phone devices. Target mobile devices are acoustically passive. They
listen to beacons emitted from anchor nodes and transmit the timestamps of the
decoded acoustic beacons to a location server wirelessly. The location
server communicates with anchor nodes to compute the locations of the
target devices. Previously, we proposed an asynchronous TDoA-based solution that uses asynchronous beacons, where anchors can transmit in an uncoordinated matter in \cite{wang16euc}. A proof-of-concept evaluation has been done using mobile phones as both anchors and target devices in a $1m\times 1m$ testbed that shows the promises of the approach. However, the transmission power limitation and microphone facing of mobile phones limit the applicability of the approach in large and complex indoor environments. When operating in large indoor areas, the effects of multi-path propagation and low received signal-to-noise ratio (SNR) are not negligible. Non-Line-of-Sight (NLOS) components may lead to larger range estimations, while low SNR results in mis-detection of preamble waveforms. Though both are well known problems in time-domain localization solutions, asynchronous beacons pose additional challenges as the pair-wise measurements between anchor nodes are also subject to both effects. 

In this paper, we improve upon the work in \cite{wang16euc} and develop ARABIS, a robust and low-cost acoustic IPS for mobile devices. Our main contributions are two-fold. First, we develop an extensible acoustic board featuring on-board power amplifier, microphone and speaker. The components are chosen to have an operational range of at least 11 meters when two anchors are facing 45 degrees away from each other and at least 50 meters when they face directly to each other. The range can be further extended by attaching external speakers. Second, to mitigate the effects of low SNR and multi-path, we devise a robust algorithm that iteratively removes possible outliers by taking advantage of redundant TDoA estimates. Experiments have been carried in two testbeds of sizes $10.67m \times 7.76m$ and $15m\times 15m$, one in an academic building on McMaster University campus and one in a convention center in Pittsburgh. The proposed system achieves average and 95\% quantile localization errors of 7.4cm and 16.0cm in the first testbed with 8 anchor nodes and average and 95\% quantile localization errors of 20.4cm and 40.0cm in the second testbed with 4 anchor nodes only. 

The rest of the paper is organized as follows. 
An overview of IPS using asynchronous acoustic beacons is given in Section~\ref{sect:asynchronous}. The hardware and algorithm design of ARABIS are presented in Sections~\ref{sect:hardware} and \ref{sect:algo}, respectively. Testbed setup and experimental evaluations are presented in Section~\ref{sect:eval}. Finally, we conclude the paper in Section~\ref{sect:conclusion}.

%

\section{Acoustic IPS using Asynchronous Beacons}
\label{sect:asynchronous}
In this section, we provide an overview of acoustic IPS using asynchronous beacons to help readers understand the rest of the work. More details can be found in \cite{wang16euc}.  

\begin{figure}[tbp]
  \centering
  \includegraphics[width=0.45\textwidth]{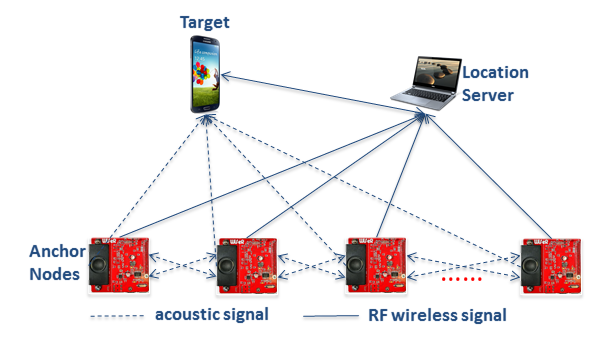}
  \caption{System architecture.}
  \label{fig:system_architecture}
\end{figure}

\paragraph*{System architecture}  
There are three types of devices in the system (see Figure~\ref{fig:system_architecture}), namely,  a
location server, anchor nodes, and target mobile devices. On the infrastructure side,
multiple acoustic anchors are deployed at fixed locations in an indoor space, each
equipped with a pair of microphone and speaker. The microphone and the speaker on an
anchor are spatially separated at a known distance.  Locations of 
all anchors are known. Periodically, the
anchors transmit acoustic beacons subject to random backoffs to minimize the
chance of collision. Each beacon message includes the identifier ({\it id}) of
the transmitting anchor node and a sequence number ({\it seqno}).  Meanwhile,
an anchor node decodes the acoustic beacons it receives including ones from
itself. Along each successful decoded beacon message, a local timestamp
({\it ts}) is recorded at the time the preamble portion of the message is
detected. 

Target mobile devices passively listen to acoustic beacons and record the {\it ids} and
{\it seqnos} from the decoded beacon messages and the timestamps of the
preambles. Both anchor nodes and mobile devices can communicate with the location
server wirelessly (e.g., via WiFi, bluetooth or Zigbee). Upon receiving
sufficient $\langle id, ts, seqno\rangle$ tuples from both sets of devices, the
location server will compute the locations of the anchor nodes and the mobile
devices using the algorithms detailed in Section~\ref{sect:algo}. 

\begin{figure}[tbp]
  \centering
  \includegraphics[width=0.35\textwidth]{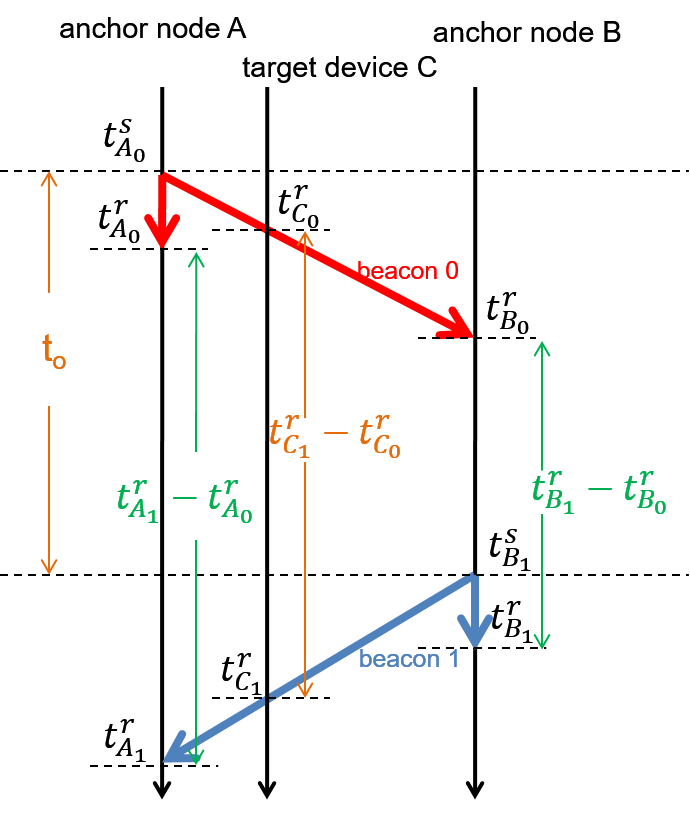}
  \caption{Determination of TDoA from a pair of anchor nodes transmitting asynchronously}
  \label{fig:timeoffset}
\end{figure}

\paragraph*{Pairwise Timeoffset and TDoA Estimation}
If beacons are transmitted at the exact same time, TDoA can be directly
measured at a target device as the time elapsed between the reception of
consecutive beacons.  Alternatively, if all anchor nodes are
synchronized~\cite{lazik2012indoor,lazik2015ultrasonic}, one can easily determine the differences in beacon
transmission times from their synchronized clocks. In our system, however, we
rely entirely on asynchronous beacons. The key idea is to utilize the full-duplex
acoustic communication capability of anchor nodes, where a node can receive and
decode its own transmitted acoustic signal.  

The TDoA estimation algorithm is best understood through an example as
illustrated in Figure~\ref{fig:timeoffset}, where there are two anchor nodes A
and B and a target device C. Consider at time $t_{A_0}^s$ that node A transmits
a beacon, which is received at time $t_{A_0}^r$, $t_{B_0}^r$, $t_{C_0}^r$ at nodes
A, B, C, respectively. All timestamps are based on local clocks. At time
$t_{B_1}^s$, anchor B transmits a beacon message, which is received at time
$t_{A_1}^r$, $t_{B_1}^r$, $t_{C_1}^r$ at nodes A, B, C, respectively. Note that
$t_{A_0}^s$ and $t_{B_1}^s$ are not known due to uncertain delays in the
acoustic interfaces.  

To compute TDoA, we need to know the interval $t_o$ between the
transmissions of the two beacons in a common reference time. Denote $d_{AA}$,
$d_{BB}$, respectively, as the distances from A's speaker to its microphone, and from B's
speaker to its microphone, respectively.
Similarly, we denote $d_{AB}$ and $d_{BA}$ respectively as
the distance from A's speaker to B's microphone and vice versa.
Let $c$ be the speed of the sound in the medium. It is easy to show that at node A,
\begin{equation}
\begin{array}{lll}
t_o  & = & (t_{A_1}^r - t_{A_0}^r) + \frac{d_{AA}}{c} - \frac{d_{BA}}{c} 
\end{array}
\label{eq:offsetA}
\end{equation} 
and, similarly at node B, 
\begin{equation}
\begin{array}{lll}
t_o  & = & (t_{B_1}^r - t_{B_0}^r) - \frac{d_{BB}}{c} + \frac{d_{AB}}{c}.
\end{array}
\label{eq:offsetB}
\end{equation}

Combining \eqref{eq:offsetA} and \eqref{eq:offsetB}, we have
\begin{equation}
\resizebox{0.45\textwidth}{!}{$
t_o = \frac{(t_{B_1}^r - t_{B_0}^r) + (t_{A_1}^r - t_{A_0}^r)}{2} + \frac{(d_{AA} - d_{BB})+(d_{AB} - d_{BA})}{2c},
$}
\label{eq:offset_original}
\end{equation}

which can be further simplified as follows under the assumptions that $d_{AB} = d_{BA}$ and $d_{AA} = d_{BB}$, 
\begin{equation}
t_o \approx \frac{(t_{B_1}^r - t_{B_0}^r) + (t_{A_1}^r - t_{A_0}^r)}{2}
\label{eq:offset}
\end{equation}

Once the transmission time offset $t_o$ is determined, the TDoA of the beacons from A and B on node C can be easily computed as 
\begin{equation}
TDoA_{A_0B_1} = t_{C_1}^r - t_{C_0}^r - t_o
\label{eq:tdoa}
\end{equation}

We also note that the distance between node A and node B could be approximated
as follows in \cite{peng2007beepbeep}:
\begin{equation}
c \times \frac{(t_{B_1}^r - t_{B_0}^r) - (t_{A_1}^r - t_{A_0}^r)}{2} + d_{AA} + d_{BB}.
\label{eq:distance}
\end{equation}

\section{Hardware Design}
\label{sect:hardware}
\begin{figure}[tbp]
  \centering
  \includegraphics[width=0.45\textwidth]{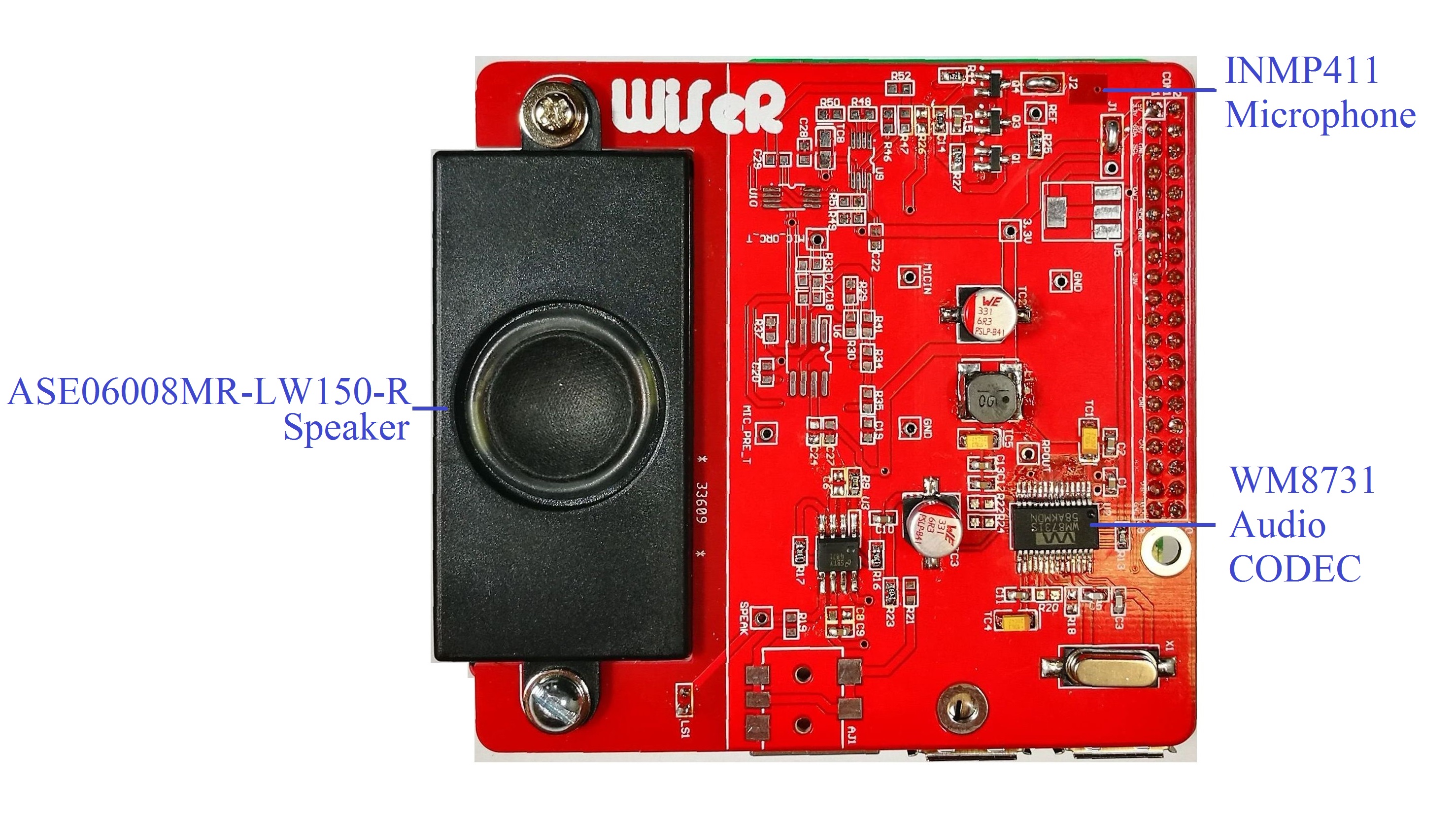}
  \caption{ARABIS Acoustic Board}
  \label{fig:acoustic_board}
\end{figure}
In ARABIS, target devices can be any smart phone equipped with a microphone that can receive acoustic beacons in the range of 17.5KHz -- 21.5KHz. Anchor nodes are purposefully designed. To allow fast prototype, we develop an acoustic board that can attach to Raspberry Pi 3 (RPI3), a single board computer that supports Linux systems (\ref{fig:acoustic_board}). In this section, we focus on the design of the acoustic board and its key components. 

\begin{figure}[tbp]
  \centering
  \includegraphics[width=0.40\textwidth]{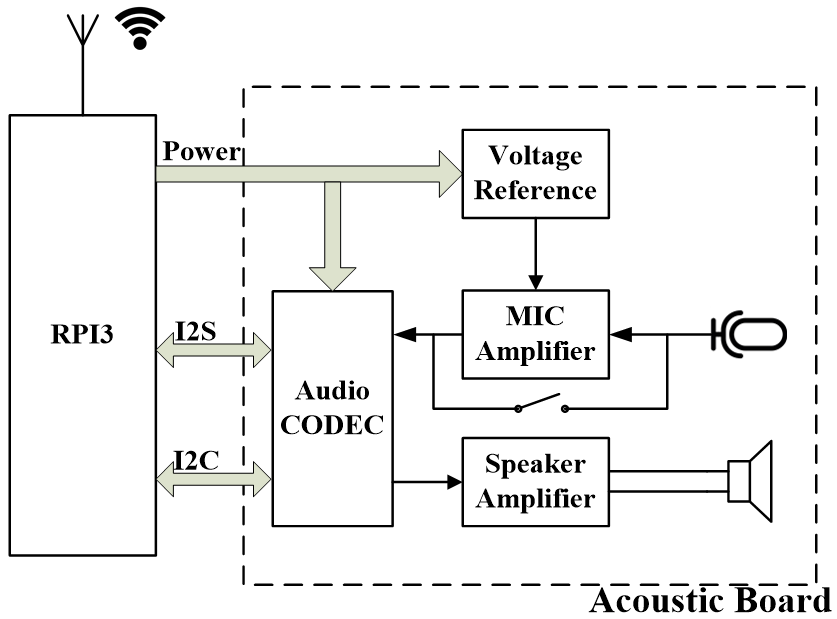}
  \caption{System block diagram.}
  \label{fig:hardware_block_diagram}
\end{figure}

The hardware diagram of the acoustic shield is shown in Figure~\ref{fig:hardware_block_diagram}. It consists of an audio CODEC IC, voltage reference, MIC amplifier circuit, speaker amplifier circuit, microphone and speaker. The board performs two functions, 1) broadcasting acoustic beacon signals, and 2) recording beacon signals both from itself and from other anchor nodes. The codec IC is controlled by RPI3 via I2C interface, and transfer to/from RPI3 the digital streams via I2S protocol.

\subsection{Codec IC}

A codec IC is a device for encoding or decoding a digital data stream. It is the kernel component that determines the sound and recording quality. WM8731 is chosen as the codec due to its superior quality, low cost and supports for I2S and I2C interfaces~\cite{WM8731}. WM8731 is a general purpose commercial codec IC widely used for portable speech players and recorders. It can support up to 96 kHz Digital to Analog conversion  (DAC) and Analog to Digital Conversion(ADC) sampling frequency, and up to 24-bit stereo. It includes microphone inputs to the on-board ADC, headphone outputs from the on-board DAC, and both of them can achieve at least 90dB SNR. Besides, when controlled via I2C, it supports volume controls, mutes, de-emphasis and extensive power management. 

\subsection{Voltage Reference}

Voltage reference provides a low noise power, stable voltage offset for the MIC Amplifier. When playing sound signals, it is crucial to minimize on-board and ambient noise, and to maximize SNR for both recording and playback. The voltage reference is introduced to reduce the on-board noises when the recording function is executed. TL431 is a programmable shunt voltage reference with guaranteed temperature stability over the entire operating temperature~\cite{TL431}. It can operate with a wide current range from 1 to 100 mA with a typical dynamic impedance of 0.22 Ω. In addition, the reference voltage tolerance can reach up to 0.5\%. 
\subsection{MIC Amplifier}
Bipolar Junction Transistor (BJT) is introduced to the MIC amplifier. The BJT is set to work in forward-active mode, where the collector-emitter current is approximately proportional to the base current. The total power of  most on-board noises is very limited with high voltage level but very small current. As a result, voltage sensitive components (e.g, ADC) can be easily affected by those noises. With the help of MIC Amplifier, we can filter part of the noises and improve SNR of incoming acoustic waveforms before feeding them to ADC. However, it shall be noted that with BJT, the amplifier gain is fixed. 
\subsection{Microphone}

\begin{figure}[tbp]
  \centering
  \includegraphics[width=0.35\textwidth]{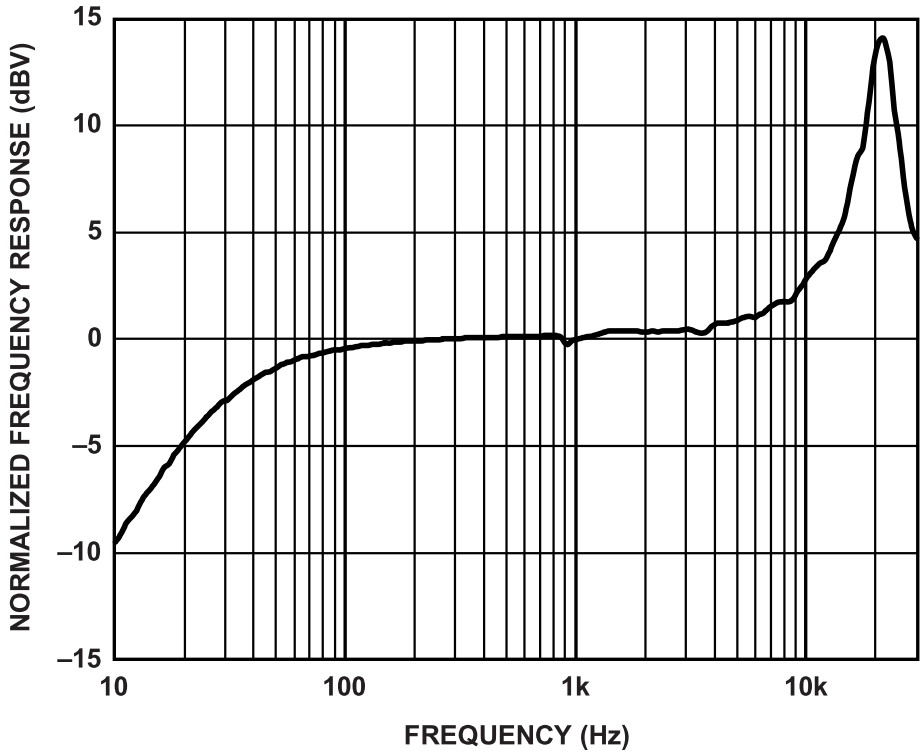}
  \caption{Typical frequency response of INMP411.}
  \label{fig:frequency_response_INMP411}
\end{figure}

To decode the beacon signal in the range of 17.5KHz -- 21.5KHz, we chose INMP411 for microphone~\cite{INMP411}. INMP411 is a high performance, high SPL, low noise, low power, analog output bottom ported, omni-directional MEMS microphone. Its sensitivity specification makes it an excellent choice for near-field applications. The INMP411 has a linear response up to 131 dB SPL. Its normalized frequency response curve (Figure~\ref{fig:frequency_response_INMP411}) shows the gain  reaches nearly 15dBV around 20 kHz comparing to 0 dBV in the range of 100 Hz to 3 kHz. Therefore, background audible sounds would not cause too much interference to beacon signals. 

\subsection{Speaker Amplifier and Speaker}
A general purpose audio amplifier IC is chosen to build the speaker amplifier. LM4871 is a mono bridged audio power amplifier capable of delivering 3W of continuous average power into a $3\Omega$ load with less than 10\% total harmonic distortion when powered by a 5V power supply~\cite{LM4871}. It provides high power, high fidelity audio output, unity-gain stability, and does not require output coupling capacitor, bootstrap capacitor or snubber network. It is ideally suited for low-power portable system.

The selection of speaker is limited by the design requirement. ASE06008MR-LW150-R is chosen for its high output power and uniform gain in the operational frequency range~\cite{ASE06008MR}. It is a 60mm enclosed speaker with 3W continuous, 4 maximum power handling and $8\Omega$ impedance. In addition, its frequency response performance in the range of 17 kHz to 20 kHz is flat and can almost reach the peak of 90 dB.
\section{Robust TDoA Trilateration}
\label{sect:algo}
The main challenge in TDoA trilateration is the degradation of timing accuracy due to low received SNR and/or existence of multiple paths. The problem is compound by the fact that the time offset approximation in \eqref{eq:offset} requires the estimation of arrival times on the anchor nodes as well. Consider a pair of anchor nodes $A$ and $B$ and target node $C$, as evident from Figure~\ref{fig:timeoffset}, the pair-wise TDoA error as the result of beacon 0 from node $A$ and beacon 1 from node $B$ arises from two sources\footnote{We assume the timestamps of the local transmissions $t^r_{A_0}$ and  $t^r_{B_1}$ are accurate due to high SNR and strong LOS component}, namely, i) inaccurate timestamps $t^r_{B_0}$ or $t^r_{A_1}$, and ii) inaccurate timestamps $t_{C_0}^r$ or $t_{C_1}^r$. 

Our proposed robust TDoA trilateration scheme is motivated by two observations. First, the bi-directional acoustic channels between two anchor nodes tend to be symmetric. As a result, inaccurate timestamps at the anchors typically imply larger pair-wise ranging errors between the anchors. Second, under proper deployment, majority of anchor nodes that are in the range of a target device should have LOS paths. Therefore, for both sources of errors, if we can remove the outlier TDoA estimates, trilateration using the remaining TDoA would be more accurate. Next, we will present the detailed procedure for outlier removal. 
\paragraph*{Detection of Outlier Time Offset}
For symmetric acoustic channels, $t^r_{B_0}$ and $t^r_{A_1}$ would error on the same side. From \eqref{eq:distance}, we see that if $t^r_{B_0}$ and $t^r_{A_1}$ change by the same amount, the distance estimate between $A$ and $B$ would differ by twice the amount divided by the speed of sound in the medium. Since anchor locations are known, by comparing the ground truth distance and the estimated distance, we can eliminate time offset estimates from the pairs with large ranging errors. For subsequent discussion, we only consider anchor pairs with valid time offset estimates.

For $N$ beacons from different anchor nodes, there are at most $N$ sets of time offset estimate pairs. If there are still multiple sets available after removing all outlier time offsets, we select the best set, which has the smallest average of pairwise ranging errors.
\paragraph*{Iterative Outlier TDoA Removal}
Under the assumptions that majority of anchors have strong direct LOS paths to the target device, we aim to eliminate timestamps from ``bad" anchor nodes. Outlier detection  (also anomaly detection) is the identification of observations which do not conform to an expected pattern or other observations in a dataset. Let $A_1, A_2, \ldots, A_m$ be the set of anchor nodes whose beacons are received by the target device.  Let $TDoA_{ij}$ be the TDoA estimate between anchors $A_i$ and $A_j$. For ease of presentation, we only consider one beacon from each anchor node, and all the time offsets are valid. The unknown target location $x$ can be determined by solving the following non-linear optimization problem:
\begin{equation} \label{eq:optimization}
 \displaystyle \min \sum_{i,j=1, i\neq j}^{m} \{c\times TDoA_{ij} - (dist(A_i, x) - dist(A_j,x))\}^2,
\end{equation}
where $dist(\cdot)$ is the Euclidean distance. The iterative Gauss-Newton algorithm~\cite{bjorck1996numerical} can be used to solve this problem.

Let $\hat{x}$ be the estimated location from \eqref{eq:optimization}. If $x$ is known, $dist(A_i, x) - dist(A_j,x))$ can be computed from the known locations of anchors $A_i, A_j$ and $x$. If the measured $TDoA_{ij}$ defers from this quantity by more than a threshold value, this implies that at least one of the beacon timestamps is erroneous. Not knowing $x$, we can use $\hat{x}$ as its approximation. Enumerating through all valid pairs, we count the number of times an anchor node contributes to ``erroneous'' TDoA estimate and remove the anchor node with the highest count from the list. This procedure repeats with the remaining anchors until only 3 (4) anchors remain for 2D (3D) localization or all TDoA estimates have small errors. Clearly, such a procedure is not guaranteed to find all outliers. However, our experimental study shows that it can indeed improve the localization accuracy. 

The pseudocode of the proposed robust TDoA trilateration algorithm is given in Algorithm~\ref{algo:outlier_timeoffset} and Algorithm~\ref{algo:outlier_tdoa}.

\begin{algorithm}
  \caption{Detection of Outlier Time Offset}\label{algo:outlier_timeoffset}
  \begin{algorithmic}[1]
    \Procedure{TimeOffsetEstimates}{$\{t^r_{i_j}\}$} \Comment{$\forall i,j \in$ the same set of anchors specified}
    \State $avgErrD \gets \infty$
    \State $r \gets NaN$ \Comment{best reference anchor ID}
    \ForAll{$i \in anchor IDs$} \Comment{as time reference}
      \ForAll{$j \in anchor IDs$}
        \State $\hat{D}_{ij} \gets Eq~\eqref{eq:distance}$ \Comment{ranging}
        \State $\hat{to}_{ij} \gets Eq~\eqref{eq:offset}$ \Comment{time offset}
        \State $errD_{ij} \gets abs(\hat{D}_{ij} - D_{ij})$ \Comment{$D_{ij}$ is ground truth}
        \If {$errD_{ij} > rangingErrThr$}
          \State $errD_{ij} \gets NaN$
          \State $\hat{to}_{ij} \gets NaN$
        \EndIf
      \EndFor
      \If {$length(\{errD_{ij}\}) \geq numAnchorsReq$} \footnotemark
        \If {$average(\{errD_{ij}\}<avgErrD$}
          \State $r \gets i$
        \EndIf
      \EndIf
    \EndFor
    \State \Comment{return a set of time offset estimate pairs}
    \If {$r \ne NaN$}
      \State \textbf{return} $\{\hat{to}_{rk}\}$ \Comment{$\{k\} \gets \{j\} \setminus \{NaN\ entries\}$}
    \Else
      \State \textbf{return} $\emptyset$
    \EndIf
    \EndProcedure
  \end{algorithmic}
\end{algorithm}
\footnotetext{A. $numAnchorsReq$ is 3(4) for 2D(3D) localization B. $NaN$ element doesn't count in $length()$ and $average()$}

\begin{algorithm}
  \caption{Iterative Outlier TDoA Removal}\label{algo:outlier_tdoa}
  \begin{algorithmic}[1]
    \Procedure{LocationEstimate}{$\{t^r_{C_i}\}$} \Comment{$C$  is  target, $\forall$ $i \in A$, where $A$ is the set of anchors from which beacons are received at $C$}
    
    \State $\{\hat{to}_{rk}\} \gets Algorithm~\ref{algo:outlier_timeoffset}$
    \If {$\{\hat{to}_{rk}\} = \emptyset$}
      \State \textbf{return} $NaN$
    \EndIf
    \State $B \gets \{k\}$
    \State $\{TDoA_{rk}\} \gets Eq~\eqref{eq:tdoa}$ \Comment{$r \in B, \forall k \in B$}
    \State $\{TDoA_{ij}\} \gets Permute(\{TDoA_{rk}\})\footnotemark$ \Comment{$\forall i,j \in B$}
    \State $\{ddoa_{ij}\} \gets \{c \times TDoA_{ij}\}$
    
    \State $isAllPairsGood \gets False$
    \Repeat
      \If{$|B| < numAnchorsReq$}
        \State \textbf{return} $NaN$
      \EndIf
      \State $\hat{x} \gets Eq~\eqref{eq:optimization}$
      \ForAll{$i \in B$}
        \State $numBadPair_i \gets 0$
        \ForAll{$j \in B, j \ne i$}
          \State $errddoa_{ij} \gets ddoa_{ij} - (d(i, x) - d(j,x))$ \footnotemark
          \If {$errddoa_{ij} > ddoaErrThr$}
            \State $numBadPair_i \gets numBadPair_i + 1$
          \EndIf
        \EndFor
      \EndFor
      
      \State $numMax \gets 0$ ; $idx \gets NaN$
      \ForAll{$i \in B$}
        \If{$numBadPair_i > numMax$}
          \State $numMax \gets numBadPair_i$ ; $idx \gets i$
        \EndIf
      \EndFor

      \If{$idx \ne NaN$}
        \State $B \gets B \setminus idx$
      \Else
        \State $isAllPairsGood \gets True$
      \EndIf
    \Until{$isAllPairsGood$}
    
	\State \textbf{return} $\hat{x}$ \Comment{return a location estimate}
    \EndProcedure
    
    
  \end{algorithmic}
\end{algorithm}
\addtocounter{footnote}{-1}
\footnotetext{$TDoA_{ij} \gets TDoA_{rj} - TDoA_{ri}, \forall i,j \in \{k\}$}
\addtocounter{footnote}{+1}
\footnotetext{$d(\cdot)$ is the Euclidean distance}
\section{Performance Evaluation}
\label{sect:eval}
In this section, we provide performance evaluations of the proposed algorithms using two experimental testbeds. We are concerned with three aspects of the performance, namely, 1) accuracy and effectiveness of the algorithms, 2) whether there exists system bias in the localization results, and 3) impacts of the number of anchor nodes used.  
\subsection{Testbed setup}
\label{sect:testbed}
\begin{figure}[tbp]
  \centering
  \includegraphics[width=0.45\textwidth]{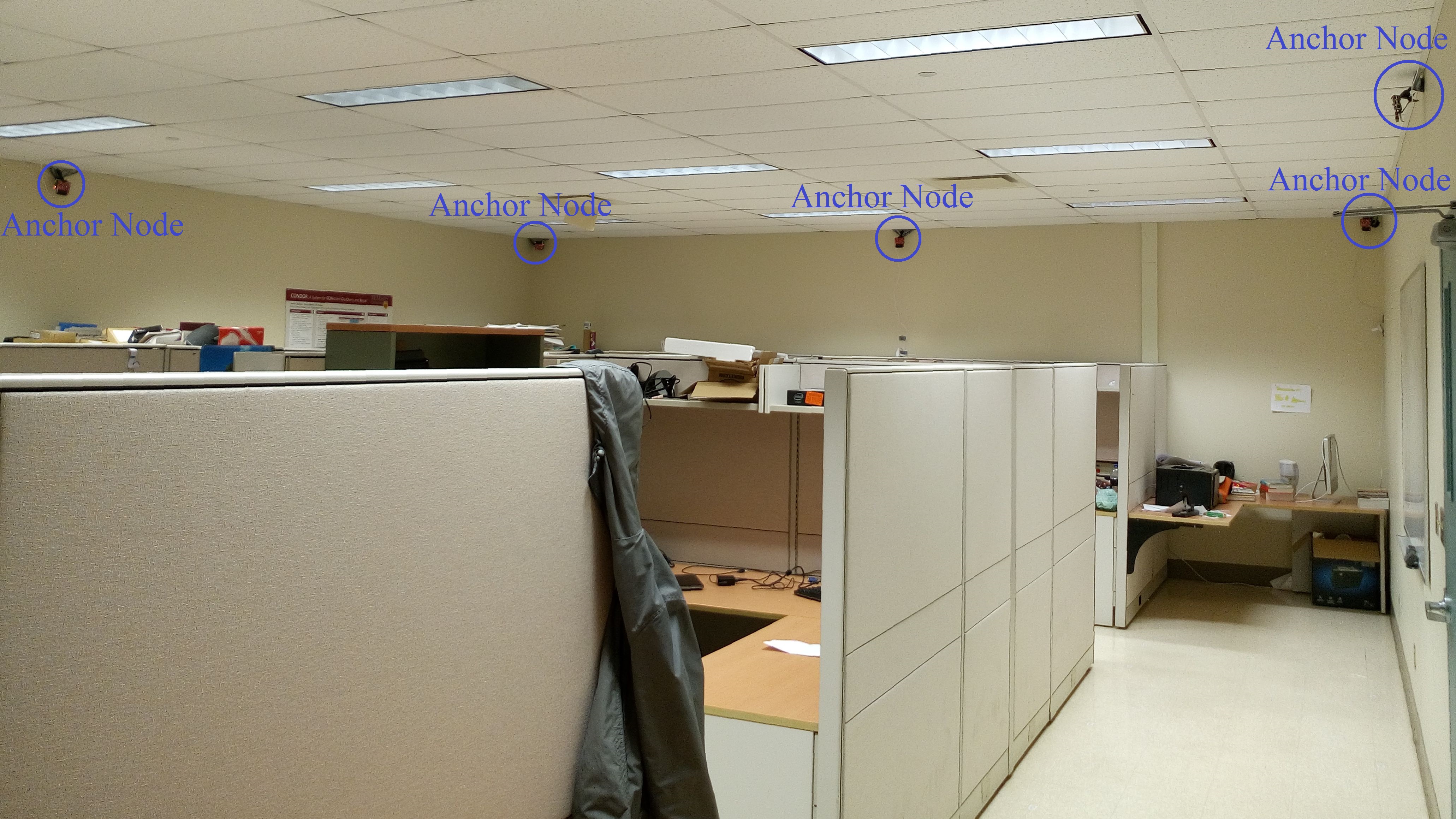}
  \caption{The first Testbed: a general crowded office environment of size ($10.67m \times 7.76m$)}
  \label{fig:testbed_itb223}
\end{figure}

The experimental testbeds consist of a location server, WiFi connectivity, and at least 4 anchor nodes.  
In the first testbed, as shown in Figure~\ref{fig:testbed_itb223}, eight anchors have been deployed near the ceiling of a $10.67m \times 7.76m$ office space. The placement of the anchors can be found in Figure~\ref{fig:scatters_AllPairs-RemoveOutlier_ITB223}. For evaluation purposes, an NTP server runs on the location server to loosely synchronize the anchor nodes. The anchor nodes take turns according to a TDMA schedule of slot length 1 second in transmitting acoustic beacons periodically. The beacon transmission time is 0.44 seconds. It should be noted that NTP time synchronization errors in local area networks are on the order of milliseconds and cannot be used for synchronous beacon transmissions. As future work, we will consider code-division-multiple-access (CDMA) to eliminate the need for coordinating beacon transmission schedules.  

\begin{figure}[tbp]
  \centering
  \includegraphics[width=0.45\textwidth]{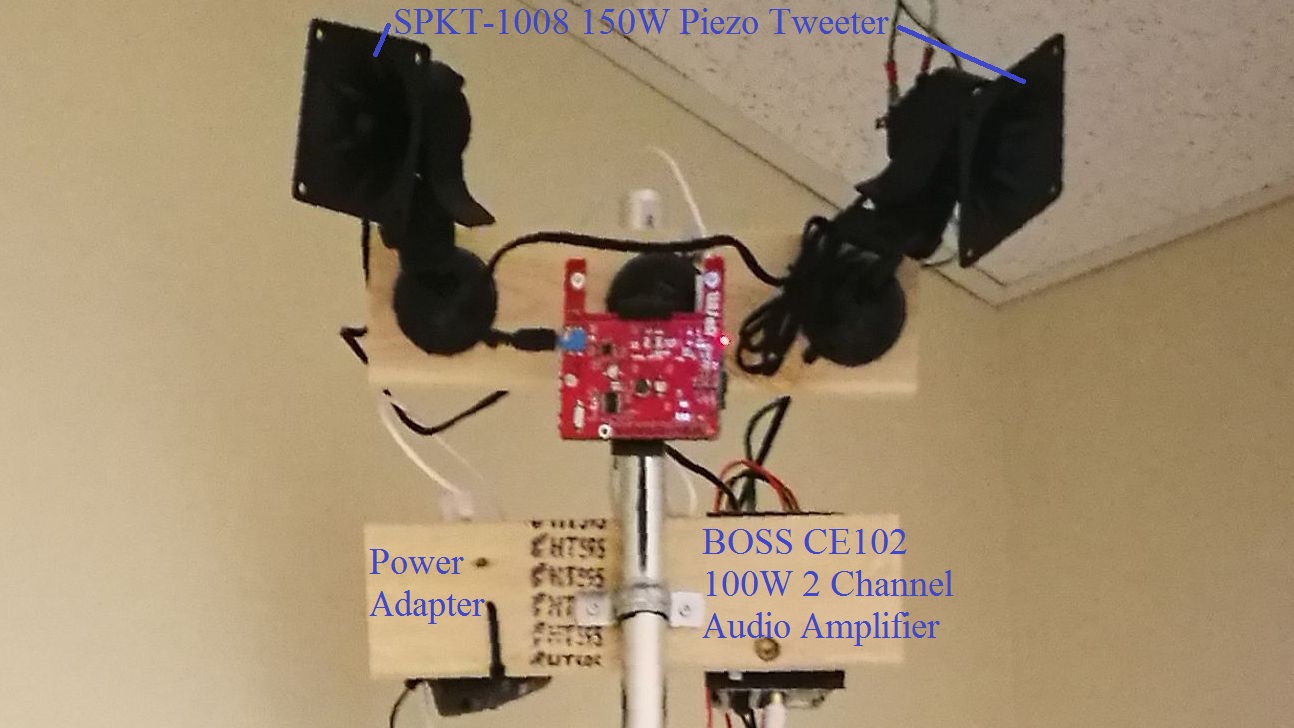}
  \caption{An anchor with external speakers used in the second testbed.}
  \label{fig:horn}
\end{figure}

\begin{figure}[tbp]
  \centering
  \includegraphics[width=0.45\textwidth]{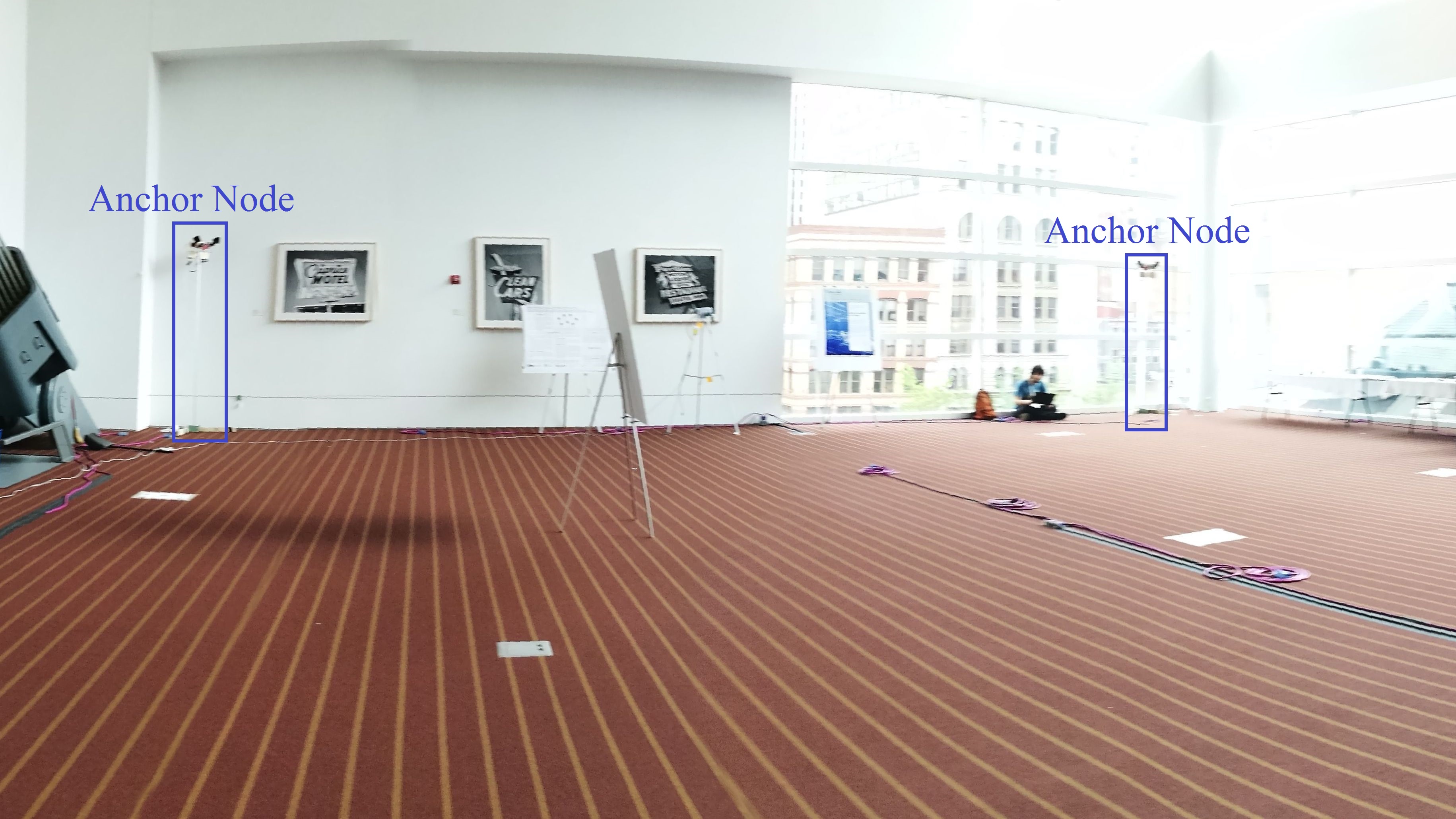}
  \caption{The second testbed: a larger and open-space environment of size about ($15m \times 15m$) in a convention center}
  \label{fig:testbed_CompetitionSite}
\end{figure}

The second testbed was of size $15m\times 15m$ and was deployed in a large convention center during the 2-day Indoor Localization Competition in the IPSN'17 conference  (Figure~\ref{fig:testbed_CompetitionSite}). As part of the requirements, only four anchors were allowed at the site~\footnote{A total of 8 anchors can be deployed but only 4 at each level of the two-floor space. Only evaluation results from the first floor are discussed as the second floor is of smaller size.}. To ensure sufficient coverage, external power amplifier and horn speakers were connected to the acoustic boards in Figure~\ref{fig:horn}. Due to limited setup time, locations of the anchor nodes were not optimized. 
Public WiFi networks were used to communicate between the location server, the target mobile device, and the anchors. In both testbeds, a Samsung Galaxy S4 phone was used as the target device. 

%
\subsection{Experiment results}
\label{sect:exp_res}
For evaluation, we have implemented 4 variations of the proposed algorithm in Python on the location server. All variations apply the detection algorithm for outlier time offset to remove invalid beacons. 
\begin{itemize}
\item Algorithm 1: All TDoA pairs, as showed in equation~\ref{eq:optimization}, without iterative outlier removal.
\item Algorithm 2: Consecutive TDoA pairs without iterative outlier removal. In this scheme, we only consider TDoA pairs from beacons that arrive in ascending time order at the target device in the time window. For instance,  let $t_1 < t_2 < t_3 < t_4$ be the timestamps of beacon messages from anchors $A_1, A_2, A_3, A_4$. We only consider TDoA estimates of $TDoA_{12}$, $TDoA_{23}$, $TDoA_{34}$, and $TDoA_{41}$.
\item Algorithm 3: All TDoA pairs with iterative outlier removal.
\item Algorithm 4: Consecutive TDoA pairs with iterative outlier removal.
\end{itemize}

One location fix is computed using beacon messages received during a time window of 18 seconds (about twice the TDMA schedule length). In theory, 4 anchor nodes are sufficient for 3D localization. However, we found that since the anchors are deployed at similar heights, errors on the vertical $z$-axis are quite large. Therefore, only $x$, $y$ coordinates are reported. In solving the optimization problem in \eqref{eq:optimization}, we confine the feasibility region to the boundaries of the testbeds. 

\begin{figure}[tbp]
  \centering
  \includegraphics[width=0.48\textwidth]{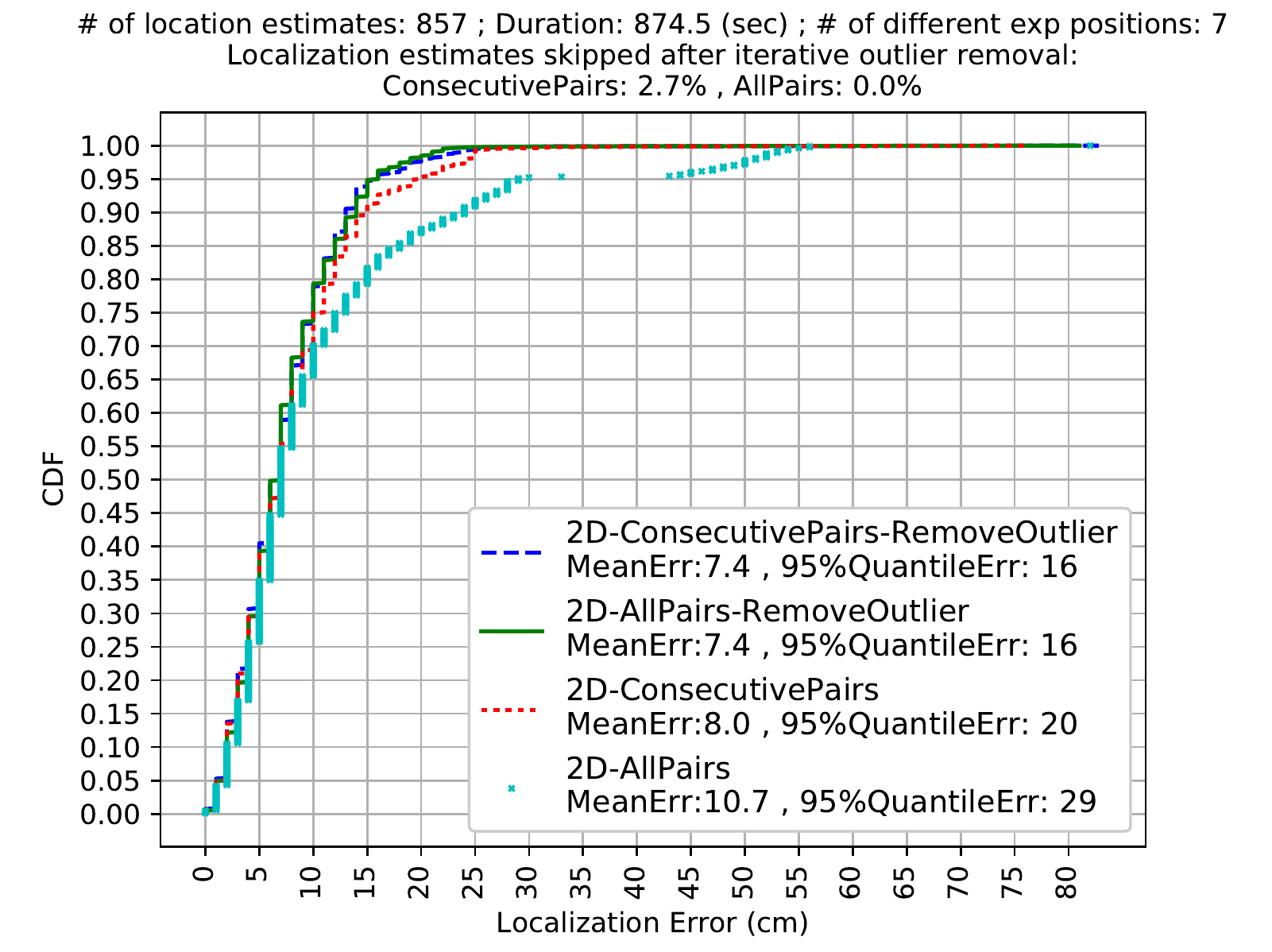}
  \caption{CDF of localization errors for different algorithms in the first testbed.}
  \label{fig:cdf_ITB223_diffAlgo}
\end{figure}

\begin{figure}[tbp]
  \centering
  \includegraphics[width=0.48\textwidth]{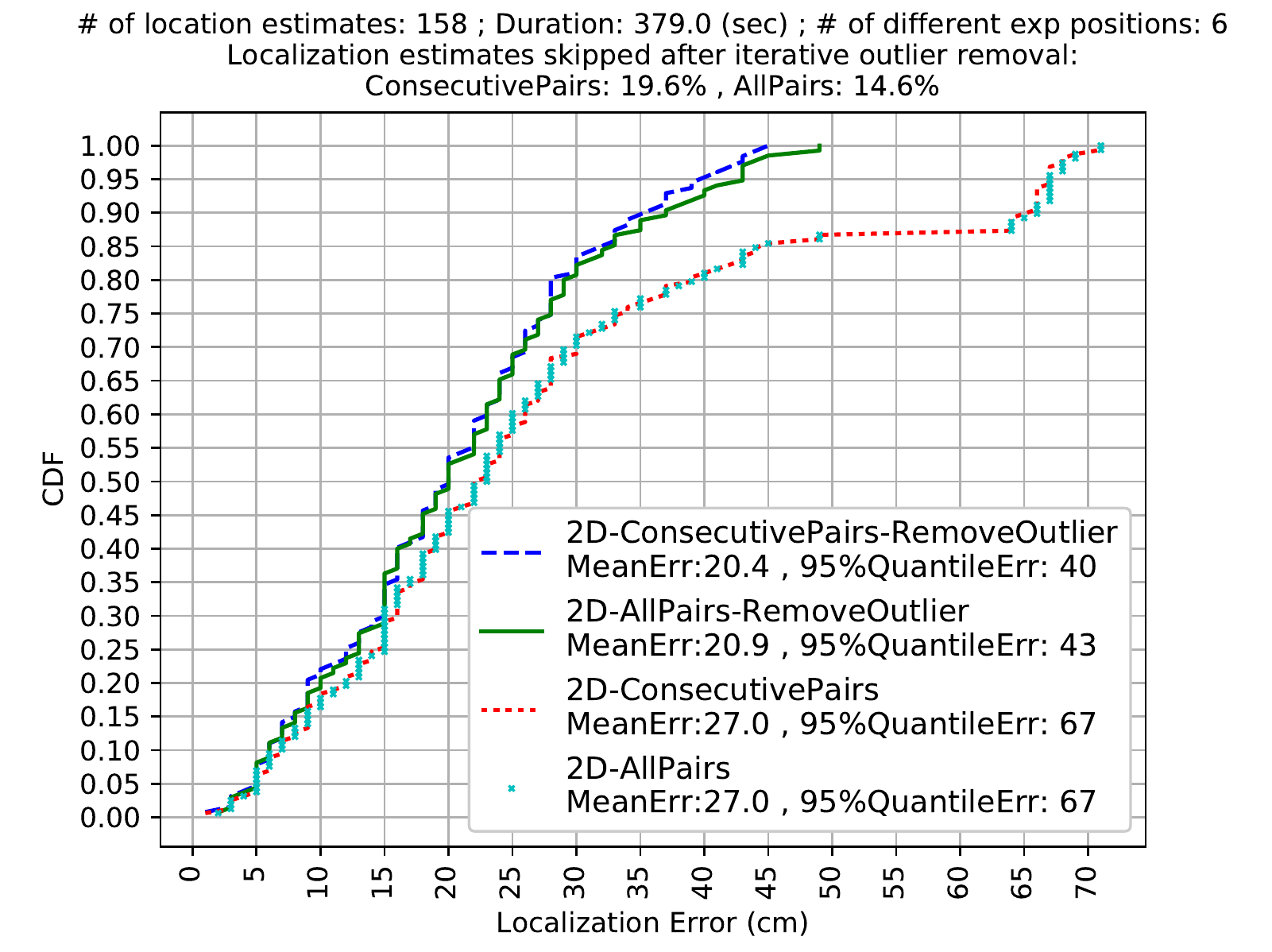}
  \caption{CDF of localization errors for different algorithms in the second testbed.}
  \label{fig:cdf_CompetitionSite_diffAlgo}
\end{figure}

Figure~\ref{fig:cdf_ITB223_diffAlgo} shows the Cumulative Density Function (CDF) of the location errors over 875 location fixes at 6 test locations in the first testbed. In this set of results, all 8 anchor nodes are used. The average localization errors for the four algorithms are 7.4cm, 7.4cm, 8.0cm and 10.7cm, respectively. The 95\% quantile localization errors are 16cm, 16cm, 20cm and 29cm, respectively. A few observations can be made from Figure~\ref{fig:cdf_ITB223_diffAlgo}. First, iterative TDoA outlier removal can indeed improve localization accuracy. Its effect is more prominent when all pairs are used since with more pairs used, chances are that some provide inaccurate TDoA estimates. In the experiment, 2.7\% of beacons are eliminated in the all-pair with outlier removal algorithm. Second, with iterative TDoA outlier removal, the performances of all-pair and consecutive-pair algorithms are comparable. This should come at little surprise. If all TDoA estimates are accurate, among eight beacons from eight anchor nodes, only seven TDoA estimates are linearly independent. Figure~\ref{fig:cdf_CompetitionSite_diffAlgo} gives the CDF of the location errors over 158 location fixes at 5 test locations in the second testbed. Similar observations can be made as in the first testbed. However, we see that the benefit of iterative TDoA outlier removal is more significant. As shown in Figure~\ref{fig:cdf_CompetitionSite_diffAlgo}, in absence of iterative TDoA outlier removal, localization errors in all-pair and consecutive-pair methods exhibit heavy tails -- both have localization errors as large as 70cm! 
From both sets of results, we observe that Algorithm 3 has the best performance. Therefore, for subsequent evaluations, only Algorithm 3 is applied. 

\begin{figure}[tbp]
  \centering
  \includegraphics[width=0.48\textwidth]{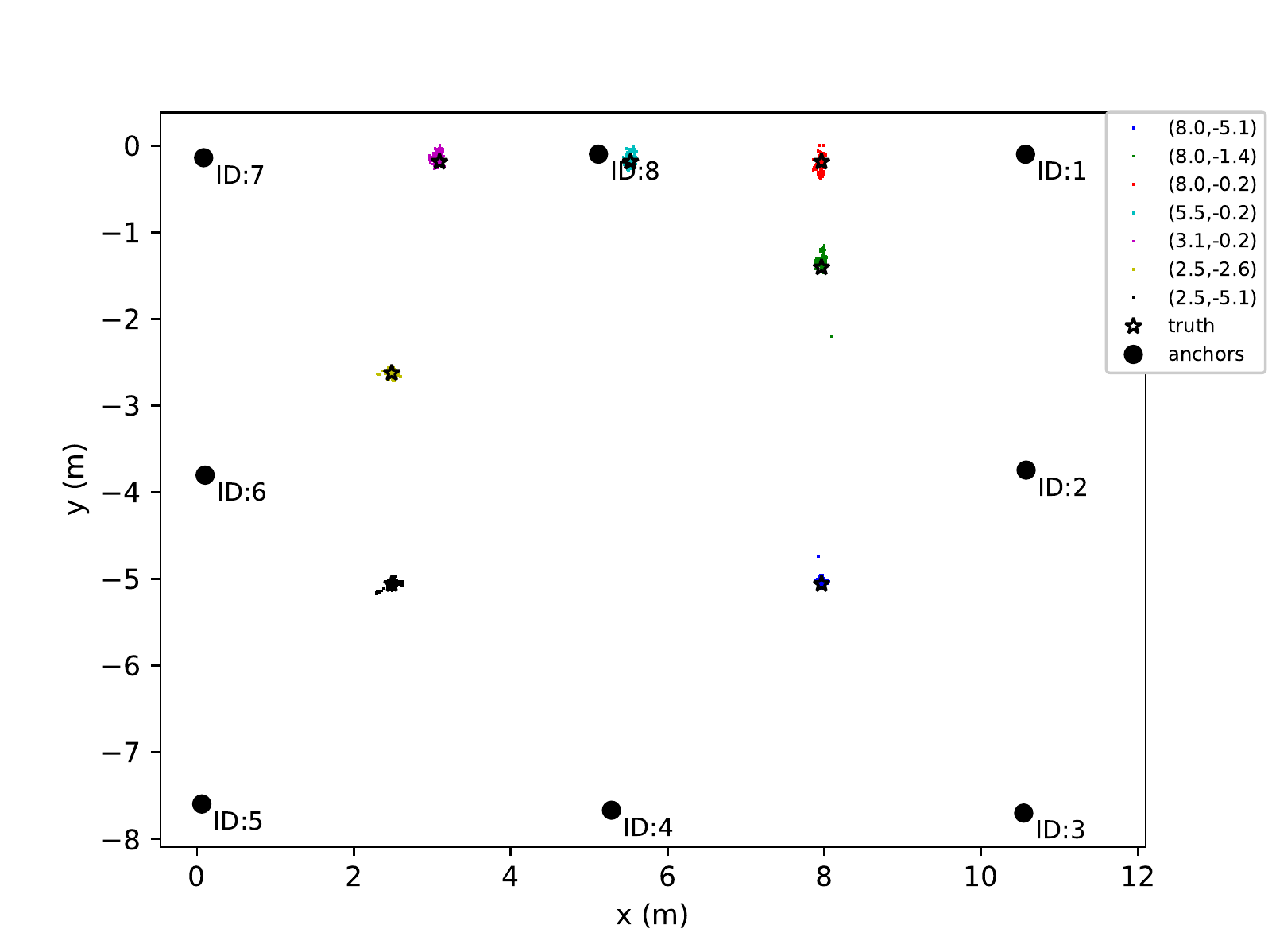}
  \caption{Scatters of localization result for algorithm 3 in the first testbed.}
  \label{fig:scatters_AllPairs-RemoveOutlier_ITB223}
\end{figure}

\begin{figure}[tbp]
  \centering
  \includegraphics[width=0.48\textwidth]{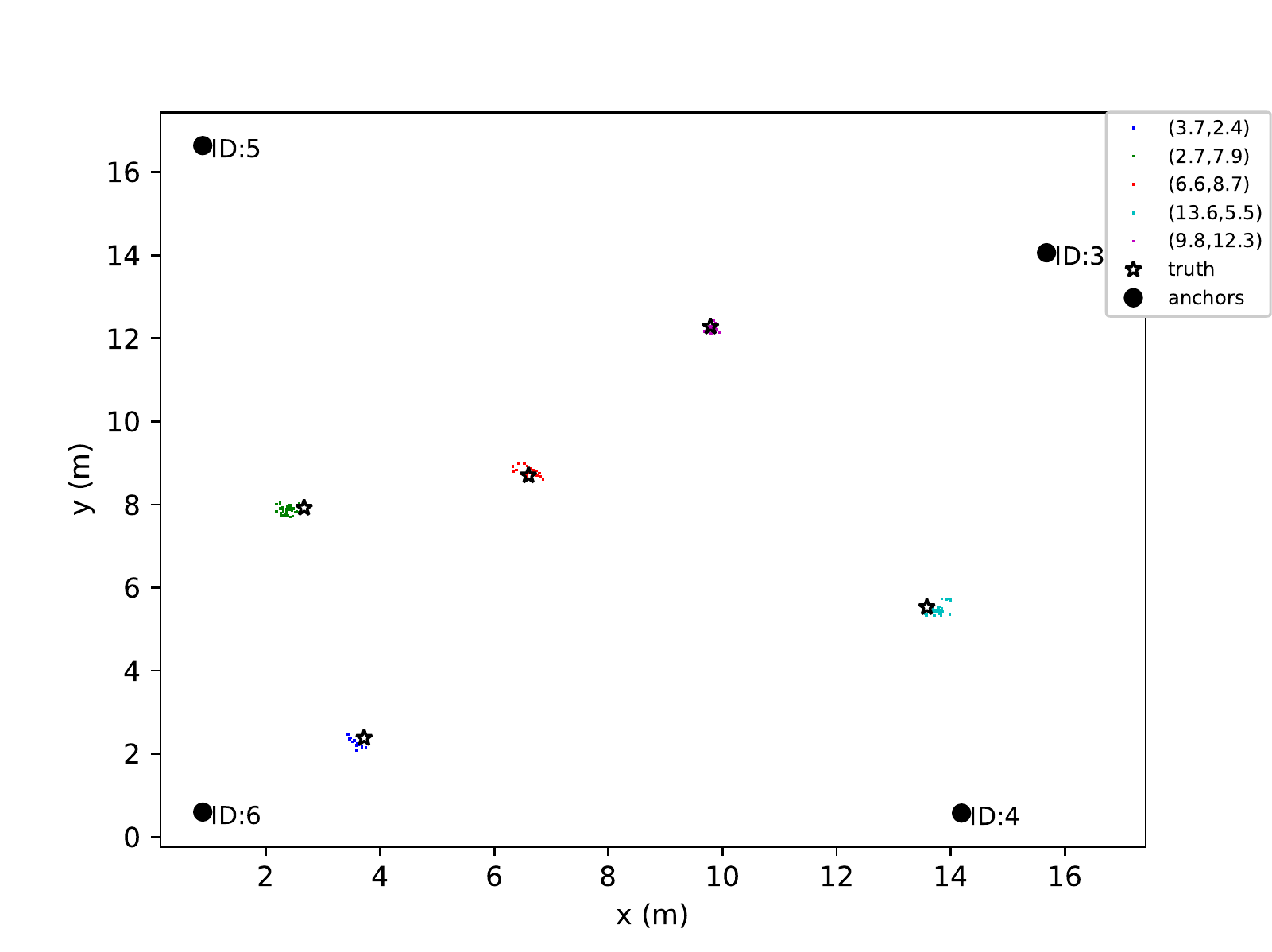}
  \caption{Scatters of localization result for algorithm 3 in the second testbed.}
  \label{fig:scatters_AllPairs-RemoveOutlier_CompetitionSite}
\end{figure}

Figure~\ref{fig:scatters_AllPairs-RemoveOutlier_ITB223} and Figure~\ref{fig:scatters_AllPairs-RemoveOutlier_CompetitionSite} are scatter plots of the localization results from the two testbed, respectively. A bias is defined as the difference between the centroid of the location estimates of one test location and the ground truth. Biases can be attributed to many reasons such as NLOS paths, errors in ground truth location, and errors in anchor locations. The average biases are respectively, 7.34cm and 28.91cm in the two testbeds. We believe the larger bias in the second testbed is likely caused by inaccurate measurement of the ground truth location, since we were holding a target device at the test locations as opposed to using a tripod as was done in the first testbed.

\begin{table}[h]
\centering
\caption{Anchor ID information.}
\label{tab:anchorIDs}
\begin{tabular}{|l|l|}
\hline
\# of anchors & anchor ID \\ \hline
4 & 2,4,6,8 \\ \hline
5 & 1,2,4,6,8 \\ \hline
6 & 1,2,3,4,6,8 \\ \hline
7 & 1,2,3,4,5,6,8 \\ \hline
8 & 1,2,3,4,5,6,7,8 \\ \hline
\end{tabular}
\end{table}

\begin{figure}[tbp]
  \centering
  \includegraphics[width=0.45\textwidth]{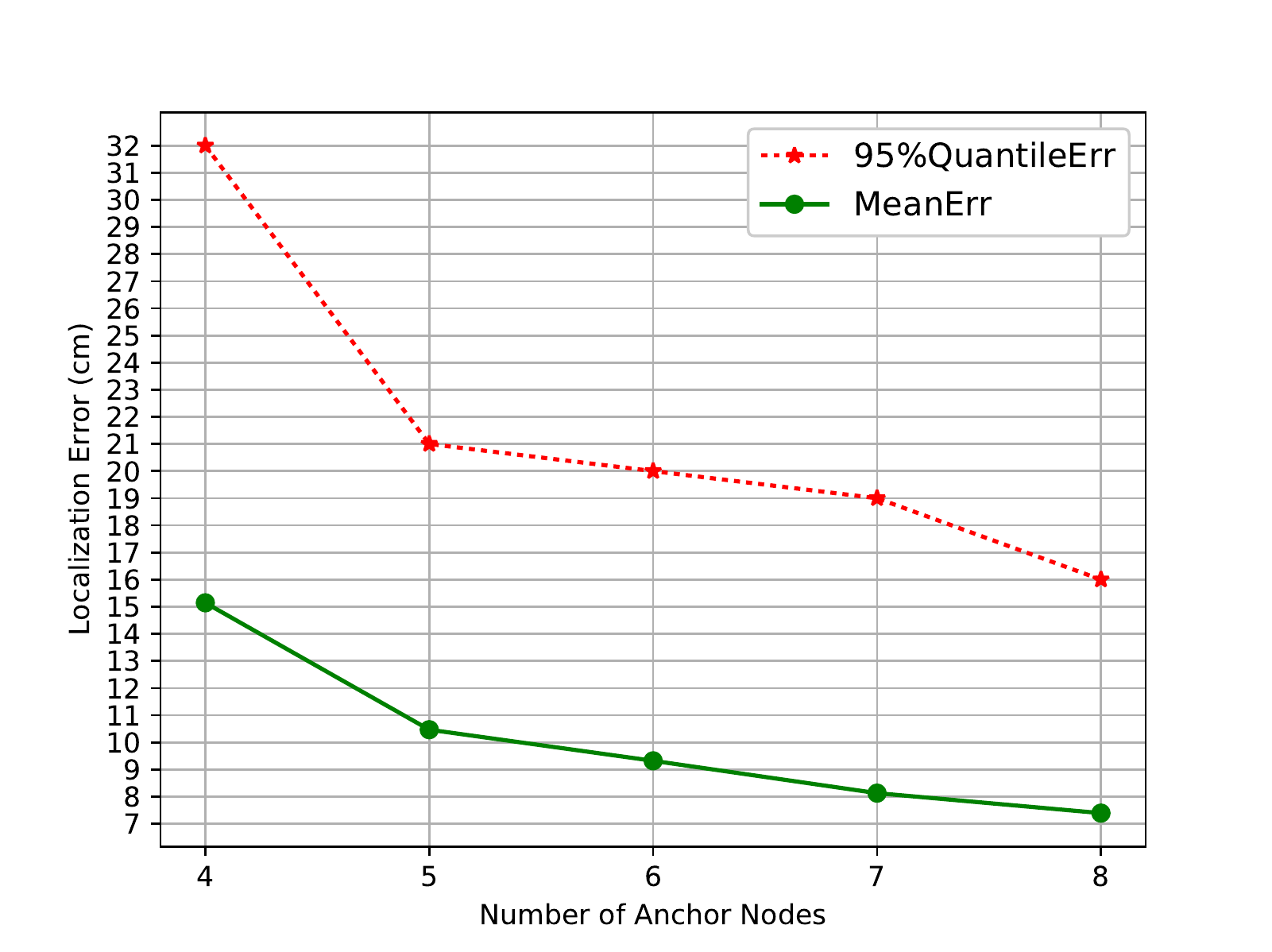}
  \caption{Localization errors for different number of anchors used in the first testbed.}
  \label{fig:locErr_ITB223_diffNumAnchors}
\end{figure}

Lastly, we evaluate the impact of the number of anchor nodes used. The dataset collected from the first testbed was used by progressively removing the anchor nodes at the four corners and their respective beacon messages. The anchor ID information is showed in Table~\ref{tab:anchorIDs}. Figure~\ref{fig:locErr_ITB223_diffNumAnchors} gives the average and 95\% quantile localization errors. As expected, the more anchors used, the better the localization accuracy is. When there are only 4 anchors in the first testbed, the localization accuracy is comparable to the one in the second testbed. Furthermore, we observe while the performance degrades gradually from 8 to 5 anchors, there is a significant drop in localization accuracy when the number of anchors decreases to 4. This may be explained by the removal of one anchor that negatively affects the localization errors of the test sites at the upper half of the area. 

\section{Conclusion}
\label{sect:conclusion}
In this paper, we proposed ARABIS, a robust and low-cost IPS using asynchronous acoustic beacons.
An extensible acoustic board is designed to support large operational ranges.
With a robust algorithm to remove outliers caused by low SNR and multi-path signals, experiments in two testbeds demonstrated the effectiveness of the proposed solution.

The limitation of the iterative outlier TDoA removal approach lies in the fact that once a target is moving, recent TDoAs in the time window become outliers and are not used until the target stops.  
As future work, we plan to devise a mechanism to handle this scenario.
\balance
\bibliographystyle{IEEEtran}
\bibliography{acoustic}

\begin{thebibliography}{10}
\providecommand{\url}[1]{#1}
\csname url@samestyle\endcsname
\providecommand{\newblock}{\relax}
\providecommand{\bibinfo}[2]{#2}
\providecommand{\BIBentrySTDinterwordspacing}{\spaceskip=0pt\relax}
\providecommand{\BIBentryALTinterwordstretchfactor}{4}
\providecommand{\BIBentryALTinterwordspacing}{\spaceskip=\fontdimen2\font plus
\BIBentryALTinterwordstretchfactor\fontdimen3\font minus
  \fontdimen4\font\relax}
\providecommand{\BIBforeignlanguage}[2]{{%
\expandafter\ifx\csname l@#1\endcsname\relax
\typeout{** WARNING: IEEEtran.bst: No hyphenation pattern has been}%
\typeout{** loaded for the language `#1'. Using the pattern for}%
\typeout{** the default language instead.}%
\else
\language=\csname l@#1\endcsname
\fi
#2}}
\providecommand{\BIBdecl}{\relax}
\BIBdecl

\bibitem{emarketer}
\BIBentryALTinterwordspacing
eMarketer. (2016) Most smartphone owners use location-based services. [Online].
  Available:
  \url{https://www.emarketer.com/Article/Most-Smartphone-Owners-Use-Location-Based-Services/1013863}
\BIBentrySTDinterwordspacing

\bibitem{mariakakis2014sail}
A.~T. Mariakakis, S.~Sen, J.~Lee, and K.-H. Kim, ``Sail: Single access
  point-based indoor localization,'' in \emph{Proceedings of the 12th annual
  international conference on Mobile systems, applications, and
  services}.\hskip 1em plus 0.5em minus 0.4em\relax ACM, 2014, pp. 315--328.

\bibitem{xiong2014synchronicity}
J.~Xiong, K.~Jamieson, and K.~Sundaresan, ``Synchronicity: Pushing the envelope
  of fine-grained localization with distributed mimo,'' in \emph{Proceedings of
  the 1st ACM Workshop on Hot Topics in Wireless}.\hskip 1em plus 0.5em minus
  0.4em\relax ACM, 2014, pp. 43--48.

\bibitem{leng2012gps}
M.~Leng, W.~P. Tay, C.~M.~S. See, and S.~G. Razul, ``Gps-free localization
  using asynchronous beacons,'' in \emph{Mobile Ad-hoc and Sensor Networks
  (MSN), 2012 Eighth International Conference on}.\hskip 1em plus 0.5em minus
  0.4em\relax IEEE, 2012, pp. 61--67.

\bibitem{peng2007beepbeep}
C.~Peng, G.~Shen, Y.~Zhang, Y.~Li, and K.~Tan, ``Beepbeep: a high accuracy
  acoustic ranging system using cots mobile devices,'' in \emph{Proceedings of
  the 5th international conference on Embedded networked sensor systems}.\hskip
  1em plus 0.5em minus 0.4em\relax ACM, 2007, pp. 1--14.

\bibitem{xu2011whistle}
B.~Xu, R.~Yu, G.~Sun, and Z.~Yang, ``Whistle: synchronization-free tdoa for
  localization,'' in \emph{Distributed Computing Systems (ICDCS), 2011 31st
  International Conference on}.\hskip 1em plus 0.5em minus 0.4em\relax IEEE,
  2011, pp. 760--769.

\bibitem{lazik2012indoor}
P.~Lazik and A.~Rowe, ``Indoor pseudo-ranging of mobile devices using
  ultrasonic chirps,'' in \emph{Proceedings of the 10th ACM Conference on
  Embedded Network Sensor Systems}.\hskip 1em plus 0.5em minus 0.4em\relax ACM,
  2012, pp. 99--112.

\bibitem{lazik2015ultrasonic}
P.~Lazik, N.~Rajagopal, B.~Sinopoli, and A.~Rowe, ``Ultrasonic time
  synchronization and ranging on smartphones,'' in \emph{Real-Time and Embedded
  Technology and Applications Symposium (RTAS), 2015 IEEE}.\hskip 1em plus
  0.5em minus 0.4em\relax IEEE, 2015, pp. 108--118.

\bibitem{wang16euc}
Y.-T. Wang, R.~Zheng, and D.~Zhao, ``Towards zero-configuration indoor
  localization using asynchronous acoustic beacons,'' in \emph{Embedded and
  Ubiquitous Computing (EUC), 2016 14th IEEE/IFIP International Conference
  on}.\hskip 1em plus 0.5em minus 0.4em\relax IEEE, 2016.

\bibitem{WM8731}
\BIBentryALTinterwordspacing
Wolfson wm8731 audio codec datasheet. [Online]. Available:
  \url{https://d3uzseaevmutz1.cloudfront.net/pubs/proDatasheet/WM8731_v4.9.pdf}
\BIBentrySTDinterwordspacing

\bibitem{TL431}
\BIBentryALTinterwordspacing
Tl431 datasheet. [Online]. Available:
  \url{http://www.ti.com/lit/ds/symlink/tl431a.pdf}
\BIBentrySTDinterwordspacing

\bibitem{INMP411}
\BIBentryALTinterwordspacing
Inmp411 datasheet. [Online]. Available:
  \url{https://www.invensense.com/wp-content/uploads/2015/02/INMP411.pdf}
\BIBentrySTDinterwordspacing

\bibitem{LM4871}
\BIBentryALTinterwordspacing
Lm4871 datasheet. [Online]. Available:
  \url{http://www.ti.com/lit/ds/symlink/lm4871.pdf}
\BIBentrySTDinterwordspacing

\bibitem{ASE06008MR}
\BIBentryALTinterwordspacing
Ase06008mr-lw150-r datasheet. [Online]. Available:
  \url{http://www.puiaudio.com/pdf/ASE06008MR-LW150-R.pdf}
\BIBentrySTDinterwordspacing

\bibitem{bjorck1996numerical}
A.~Bj{\"o}rck, \emph{Numerical methods for least squares problems}.\hskip 1em
  plus 0.5em minus 0.4em\relax Siam, 1996.

\end{thebibliography}
\end{document}